# A Deep-Learning Framework for the Generalized Emulation of Atmospheric Radiative Transfer at Microwave Frequencies

Spencer R. Jones*, Simon Pfreundschuh*, and Christian D. Kummerow*

**Department of Atmospheric Science, Colorado State University, USA*

Abstract: This paper presents an application of machine learning to the problem of the computational expense of atmospheric radiative transfer calculations from physics-based models. While in recent studies individual emulators have been created that emulate certain radiative transfer schemes for different sensors, we show that a general framework can be applied to emulate the radiative transfer method itself. A fully connected deep neural network has been trained to predict top-of-atmosphere microwave radiances in all-sky conditions over various land, ocean and sea ice surfaces at all frequencies from 1 to 300 GHz, and for all incidence angles from 0° to 65°. The reference models against which the general emulator has been trained include an Eddington two-stream radiative transfer model for one-dimensional and pseudo-three-dimensional microwave brightness temperature simulations. This neural network architecture has also been separately trained to reproduce the explicit calculations of the Community Radiative Transfer Model (CRTM) for a single microwave imager. In all cases, the model architecture is successfully able predict microwave brightness temperatures by being trained directly on the optical properties of the atmosphere and surface and therefore directly emulating the radiative transfer method. Compared to the explicit schemes used to produce the training data, the GPU-enabled emulator is able to produce brightness temperature simulations more than three orders of magnitude faster and with a root-mean-squared-error less than 0.0348 K, 0.0531 K, and 0.0246 K against the independent test data for the two Eddington models and CRTM respectively. Additionally, preliminary training using a full 3D Monte Carlo scheme for high-resolution simulations shows promising results for the application of this method to “exact” radiative transfer methods, and these results are discussed in Section IV. A pretrained version of this tool and instructions for training against a new radiative transfer scheme are available as free software at the link provided in the Data Availability section of this article.

## I. Introduction

Many satellite remote sensing techniques are based on algorithms that require atmospheric radiative transfer simulations in order to invert the measured radiances emerging from the top of the atmosphere into the geophysical scene that produced them. Minimum Variance Solutions (Duncan et al., 2018; C. D. Kummerow, 2020) and Optimal Estimation (Boukabara et al., 2011; Rodgers, 2000; Schulte & Kummerow, 2019) methods require multiple radiative transfer simulations per observed data point, and often with millions of data points per orbit, these simulations can quickly become the operational bottleneck for any scientific data application. Even Bayesian methods (C. D. Kummerow et al., 2015; Pfreundschuh et al., 2023), and full machine learning retrievals (Pfreundschuh et al., 2022, 2026), also need multiple millions of simulations in order to produce robust training data volumes. Schemes to assimilate satellite data into numerical weather models also require fast and accurate radiative transfer simulations in order to map observations to the state vector space in variational approaches (de Rosnay et al., 2022; Zhu et al., 2016).

This has led to many approximations intended to increase the speed of radiative transfer models needed to ensure operational capability for satellite retrievals (Boukabara et al., 2011) and data assimilation (Liu & Boukabara, 2014; R. Saunders et al., 1999). In the microwave spectrum, assumptions such as single scattering, two-stream methods, and the one-dimensional (1D) assumption, i.e. that the atmosphere can be approximated to only vary in the vertical direction and is homogeneous in the horizontal, have been applied ubiquitously in an effort to dramatically reduce the computational cost of these models (Grecu et al., 2016; Johnson et al., 2023; Jones & Kummerow, 2025; R. Saunders et al., 1999; Roger Saunders et al., 2018).

Recent applications of deep learning techniques to this problem have shown promise for the emulation of physics-based radiative transfer. The main improvements are the very fast speeds and reduction in computational cost gained by invoking neural-network-based solutions (Ukkonen, 2022). These machine-learning models are trained to regress the highly nonlinear relationships between geophysical parameters that describe the state of the atmosphere and surface and the simulated radiances from the physics-based radiative transfer scheme. On the numerical weather and climate modeling front, a few applications of deep learning to represent subgrid-scale physics have been successful in reducing the computational cost of parameterizations (Gentine et al., 2018; Kochkov et al., 2024; Rasp et al., 2018), including that of radiative transfer for energy fluxes (Ukkonen, 2022).

In addition to experiments with RT emulation to reduce the computational cost for models, a few recent examples of emulating fully physical RT models for observational applications have also found success (Howard et al., 2026; Liang et al., 2022). In these studies, FNNs were trained to emulate TOA radiances calculated by the Community Radiative Transfer Model (CRTM; Johnson et al., 2023), used operationally at the Joint Center for Satellite Data Assimilation. However, since CRTM contains parameterizations for each sensor, this approach is limited only to the sensors in the CRTM library and only to assumptions made by CRTM with regards to hydrometeor distributions and surface properties. Therefore, using this type of approach to simulate observations from future proposed or hypothetical sensor configurations remains impossible, despite the fact that the greatly reduced computational cost by such a model would likely have many applications.

This study explores the application of deep learning to the core radiative transfer method itself, which instead of focusing on emulating an entire radiative transfer code suite by predicting from the geophysical parameters, the TOA radiances are predicted directly from the atmospheric optical properties—namely, these are the temperature profile, extinction and scattering parameters, and surface emission characteristics. The main benefit to this approach is that the emulator becomes agnostic to microphysical and emissivity assumptions and does not have to be retrained if these schemes are changed. This allows users to get the benefit of the speed gains by machine learning prediction while also allowing them to specify their own microphysical or emissivity schemes to go from geophysical parameters to optical properties, thereby removing dependence on fully integrated, preconstructed scattering codes or lookup tables. Direct prediction from optical properties also removes the explicit dependence on frequency as a predictor, since this information is included in the calculation or lookup of optical properties.

The remainder of this paper is organized as follows: Section II describes the reference models and training data used, Section III shows training results and model performance metrics, and Section IV briefly discusses some preliminary training of the core radiative transfer emulator from a full 3D backward Monte Carlo scheme in order to estimate the computational resources and practicality of applying this emulation method to an exact radiative transfer method. Finally, Section V presents a summary and some closing discussion.

## II. Model Descriptions and Training Data

### A. Reference Models

The reference models used to generate the training data and the prediction targets are physics-based radiative transfer schemes that calculate the upwelling microwave radiances emerging from the top of a model atmospheric profile for given frequencies. The model atmospheres used to create the training data are taken from global ECMWF ReAnalysis 5 (ERA5; Hersbach et al., 2020) fields at 27 vertical pressure levels from 1000 to 100 hPa, described further in Section II B.

Three separate radiative transfer models were used to illustrate the capabilities of the emulation framework. Each model is used to generate simulated microwave radiances in brightness temperature ($T_b$) space, with each of them assuming various degrees of horizontal structure within the model atmosphere. Two models, one based on the two-stream Eddington Approximation (C. Kummerow, 1993) and the Community Radiative Transfer Model (Johnson et al., 2023), assume a plane-parallel atmosphere with homogeneous extinction and scattering properties within each layer that extend infinitely in both horizontal directions—also known as the one-dimensional (1D) assumption. Beam geometry is handled via a simple path-length adjustment through each layer as a factor of $1/\cos(\theta)$, where $\theta$ is the incidence angle of the radiation. A pseudo-three-dimensional (p3D) model also based on the Eddington Approximation (Olson et al., 2001) allows for the separate treatment of the upwelling and downwelling columns of radiation, which is appropriate whenever the scene is being viewed at an angle away from nadir. Hereafter, 1DEd, CRTM, and p3DEd will be used to refer to the one-dimensional, plane-parallel Eddington model, the Community Radiative Transfer Model, and the pseudo-3D Eddington model, respectively. The physical approach of each of these schemes are well established in the cited literature and will therefore not be extensively reviewed here, but a brief overview of the aspects relevant to the creation of the emulator training data is warranted.

1. the Eddington 1D Model

The Eddington approximation for microwave frequencies of the full two-stream solution for upwelling and downwelling radiances (C. Kummerow, 1993) consists of expanding them using Legendre functions and truncating after the first two expansion terms:

$$I(z, \theta, \phi) \approx I_0(z) + I_1(z) \cos \theta \tag{1}$$

where $I$ is the full 3D radiance function in spherical coordinates, z is the geometric height above the surface, $\theta$ is the zenith angle, $\phi$ is the azimuth angle. It can be then shown that this leads to two first-order differential equations for the isotropic and linear terms $I_0$ and $I_1$:

$$\frac{dI_0(z)}{dz} = -k(z)[1 - a(z)g(z)]I_1(z) \tag{2a}$$

$$\frac{dI_1(z)}{dz} = -3k(z)[1 - a(z)][I_0(z) - T(z)] \tag{2b}$$

where $k$ is the volumetric extinction coefficient, $a$ is the single scatter albedo, $g$ is the asymmetry parameter, and $T$ is the temperature profile.

In the Eddington models used to generate the training data, the radiances are calculated numerically, so that the atmosphere is modeled as existing of discrete layers, and values of temperature and geometric height are assigned to the layer boundaries while $k$, $a$, and $g$ are assigned to the layers and are considered homogeneous within the layer. Additional information needed involves the source term, which is the surface emission, and this is simply a function of surface emissivity and temperature. Therefore, the set of optical parameters for Eddington calculations is the following: the height profile of each layer boundary above the surface $z$, the layer boundary temperature profile $T$, the extinction coefficient $k$, the single scatter albedo $a$, the asymmetry parameter $g$, the zenith angle $\theta$, and the surface temperature $T_s$ and emissivity $\epsilon$. These parameters are also used as inputs to the emulator. Additionally, the hemispheric average emissivity $\bar{\epsilon}$ was also used as an input since it is related to the upwelling stream boundary conditions of the Eddington approximation (see C. Kummerow, 1993).

2. The Eddington p3D Model

The p3DEd approach is simply an innovation on the 1DEd approach that gives greater accuracy if the scene is being viewed at an angle away from nadir, since the downwelling and upwelling columns of radiation are treated separately. In this approach, diffuse radiation is first calculated downward where the source term is the cosmic microwave background temperature (nominally $T_{b,cosmic} = 2.7$ K) and atmospheric scattering and emission sources contribute to the radiance incident upon the surface. Then, the reflected portion of this incident radiation is calculated along with the surface emission, and this becomes the source term for the upwelling stream of radiation, which undergoes scattering and emission interactions along the upwelling column before emerging at the TOA. This is called the pseudo-3D approach because it allows the upwelling and downwelling streams to interact with scattering and emission sources that may appear in one direction of radiation stream travel and not in the other. Although this approach is not considered an "exact" handling of the full 3D RT solution as is a Monte Carlo or other 3D solutions, it offers some improvement over the 1D assumption for oblique view angles (Olson et al., 2001). Furthermore, it stands to reason that if there is no scattering source in the path of the direct stream and surface reflection can be considered specular, the p3D solution becomes equivalent to the full 3D solution.

3. the Community Radiative Transfer Model (CRTM)

This model was chosen to emulate due to its extensive use in data assimilation schemes, and therefore to show that it is also possible to apply the generalized emulation method to a highly parameterized RT model that has been extensively developed to accurately produce TOA simulated radiances as they would be seen by operational sensors. For this study, we chose to emulate CRTM for the GPM Microwave Imager (GMI; Draper et al., 2015), which measures naturally emitted microwave radiation at 10.65, 18.7, 23.8, 36.5, 89.0, 166.5, 183.31±3, and 183.31±7 GHz in 13 channels at both horizontal and vertical polarizations for 10.65, 18.7, 36.5, 89.0, and 166.5 GHz and vertical polarization only for 23.8, 183.31±3, and 183.31±7 GHz. Since the CRTM suite uses the layer optical depth $\tau$ instead of the volumetric extinction coefficient $k$ for radiance calculations, we use $\tau$ as the extinction predictor for the version of the emulator trained on CRTM. This is simply calculated as $\tau = k\Delta z$ for each profile. The inputs to the emulator for all models are summarized in Table 1.

| Predictor | Description | Reference Models | Units | Data Source |
|---|---|---|---|---|
| $T$ | Temperature | All | K | ERA5 |
| $z$ | Height | 1DEd, p3DEd | km | ERA5 |
| $k$ | Extinction Coefficient | 1DEd, p3DEd | $km^{-1}$ | Calculated |
| $\tau$ | Optical Depth | CRTM | None | Calculated |
| $a$ | Single Scatter Albedo | All | None | Calculated |
| $g$ | Asymmetry Parameter | All | None | Calculated |
| $T_s$ | Surface Temperature | All | K | ERA5 |
| $\epsilon$ | Surface Emissivity | All | None | Calculated |
| $\bar{\epsilon}$ | Hemispheric Average Surface Emissivity | 1DEd, p3DEd | None | Calculated |
| $\theta$ | Earth Incidence Angle | 1DEd, p3DEd | ° | Randomly Assigned |

B. Training, Testing, and Validation Data

The goal in building the training dataset for this study was to strike a compromise between using randomness to make the model as flexible and generally trained as possible yet staying grounded in natural variability in order to avoid training on unphysical data. The training data for all emulators built for this study consisted of ECMWF ReAnalysis 5 (ERA5; Hersbach et al., 2020) atmospheric profiles, where global hourly data was subsampled to 6-hour timesteps (0000, 0600, 1200, and 1800 UTC) from the 1st and 15th day of each month of the year 2024 to capture diurnal and seasonal variability. Similarly, validation and testing data were constructed from January, April, July, and October of 2025, where the 1st of each month was used for validation, and the 10th of each month was used for testing. In total, 99,671,040 profiles were used to create the training data, 16,611,840 profiles were used for validation, and another 16,611,840 profiles were used for testing. All three datasets were fully separated from each other, and both the validation and testing data were withheld from the model optimization process during training to avoid data leakage. The validation data was also randomly subsampled by one tenth to improve training speed and also to ensure the loss against an independent dataset was being reduced during training.

For the 1DEd and p3DEd models, each profile was randomly assigned a frequency from a uniform distribution between 1 and 300 GHz and an earth incidence angle (EIA) from a uniform distribution between 0° and 65°, inclusive. A surface emissivity for each profile over ocean was calculated using the assigned frequency and EIA, surface temperature, and wind speed from ERA5. Additionally, since microwave surface emissivity at low frequencies is sensitive to sea surface salinity, each profile was also randomly assigned a salinity value from a Gaussian distribution, where the mean and standard deviation of the distribution were obtained from a year of gridded monthly Aquarius sea surface salinity retrievals (Le Vine et al., 2018). For land and sea ice, the physical model of Weng et al. (2007) was used, which estimates surface emissivity based on characteristics such as surface temperature, snow cover, and vegetation properties. Emissivity at both horizontal and vertical polarizations were used to ensure that the natural variability of this parameter was captured in the training data.

Atmospheric optical properties were calculated using the profiles of temperature, pressure, humidity, cloud liquid water, cloud ice, rain, and snow water content from the reanalysis, along with the assigned frequencies for each profile. Assuming Mie scattering by both liquid and ice particles, the optical properties sought for the predictor data were the volumetric extinction coefficient $k$, the single scatter albedo $a$, and the asymmetry parameter $g$ for each layer derived from the full scattering phase function. For water and ice clouds in the 1DEd and p3DEd models, this was done using a monodisperse distribution, where the diameter of all liquid and ice cloud particles were taken to be 10 μm and 30 μm, respectively. In order to capture maximum variability for the calculated optical properties, but also to remain within physically realistic boundaries, a normalized gamma distribution (Testud et al., 2001) was assumed for rain and snow scattering with two moveable parameters, the normalized intercept parameter $N_w$ and the mean particle diameter $D_m$, with the latter being constrained by the layer volumetric water content $C_w$.

$$n(D) = N_w f(\mu) \left(\frac{D}{D_m}\right)^{\mu} \exp\left[-(4-\mu)\left(\frac{D}{D_m}\right)\right] \tag{3}$$

$$f(\mu) = \frac{\Gamma(4)}{4^4} \frac{(4+\mu)^{4+\mu}}{\Gamma(4+\mu)} \tag{4}$$

$$D_m = \left(\frac{4^4 C_w}{\rho \pi N_w}\right)^{\frac{1}{4}} \tag{5}$$

The shape parameter $\mu$ was fixed to 2, a value taken to be valid for the natural variability of precipitation systems in Grecu et al. (2016). $N_w$ was randomly assigned using distribution parameters obtained by analyzing four months (January, April, July, and October) of CORRA retrievals of frozen and liquid hydrometeor profiles. Furthermore, since scattering of microwave radiation due to ice particles in the atmosphere is sensitive to the ice particle density (Jones & Kummerow, 2025), a random snow particle density was assigned to each profile, again from a uniform distribution between 0.05 and 0.9 g cm$^{-3}$. Table 2 summarizes the atmospheric optical property computations for both Eddington Models. For the CRTM emulator training data, the optical properties were calculated using CRTM's internal optical property absorption and scattering lookup tables, and the only moveable parameter, effective radius, was kept at default values for each cloud type.

Table 2: Assumed particle size distributions and parameter assignment used to construct the training, testing, and validation data for the Eddington 1D and pseudo-3D emulators.

| Variable | Size Distribution | Parameters |
|---|---|---|
| Cloud Water | Monodisperse | $D_{cld,l} = 10$ µm<br>$\rho_l = 1$ g cm$^3$ |
| Cloud Ice | Monodisperse | $D_{cld,i} = 30$ µm |
| Rain Water | Normalized Gamma | $\log N_{w,r} \sim \mathcal{N}(6.90, 0.09)$<br>$\mu = 2$<br>$\rho = 1$ g cm$^3$ |
| Snow Water | Normalized Gamma | $\log N_{w,s} \sim \mathcal{N}(6.99, 0.08)$<br>$\mu = 2$<br>$\rho \sim \mathcal{U}(0.05, 0.9)$ g cm$^{-3}$ |

To summarize, combining randomly assigned PSD parameters with ERA5 atmospheric profiles and surface analyses allowed for a wide range of variability in the training data. This again was done in an attempt to remove explicit dependence on any particular microphysical or surface emissivity scheme, since machine learning algorithms are sensitive to outliers and can perform poorly when predictors are out-of-sample (Rasp et al., 2018).

C. Emulator Structure

The emulator in all cases is a deep Fully connected Neural Network (FNN) architecture, with five hidden layers, each consisting of 512 fully connected neurons. After each layer except the output layer, a Gaussian Error Linear Unit (GELU; Hendrycks & Gimpel, 2016) was applied. To reduce any tendencies during training for the model to preferentially optimize against some predictors and not others due to varying dynamic ranges in the input data (Cabello-Solorzano et al., 2023), the predictors were rescaled using common normalization techniques. Since extinction is exponential in nature, the values of $k$ were normalized in logarithmic space to somewhat linearize the distribution according to

$$\hat{k} = \frac{\ln k - \overline{\ln k}}{\sigma_{\ln k}} \tag{6}$$

where $\hat{k}$ is the normalized or encoded value of the volumetric extinction coefficient $k$. For all other predictors, the inputs were normalized using min/max normalization:

$$\hat{x} = a + (b - a)\left(\frac{x - x_{min}}{x_{max} - x_{min}}\right) \tag{7}$$

Where $\hat{x}$ is the normalized value of the predictor. $a$ and $b$ are the bounds of the normalization, and these bounds were set to -3 and 3, respectively, to approximate the dynamic range of $\hat{k}$. For training, the prediction outputs are similarly encoded in normalized space, and this was done by normalizing the "truth" values of $T_b$ that were calculated by the reference models. The resulting targets vary from -3 to 3, and the predicted values must be denormalized to physical $T_b$ space after execution, as illustrated in Fig. 1.

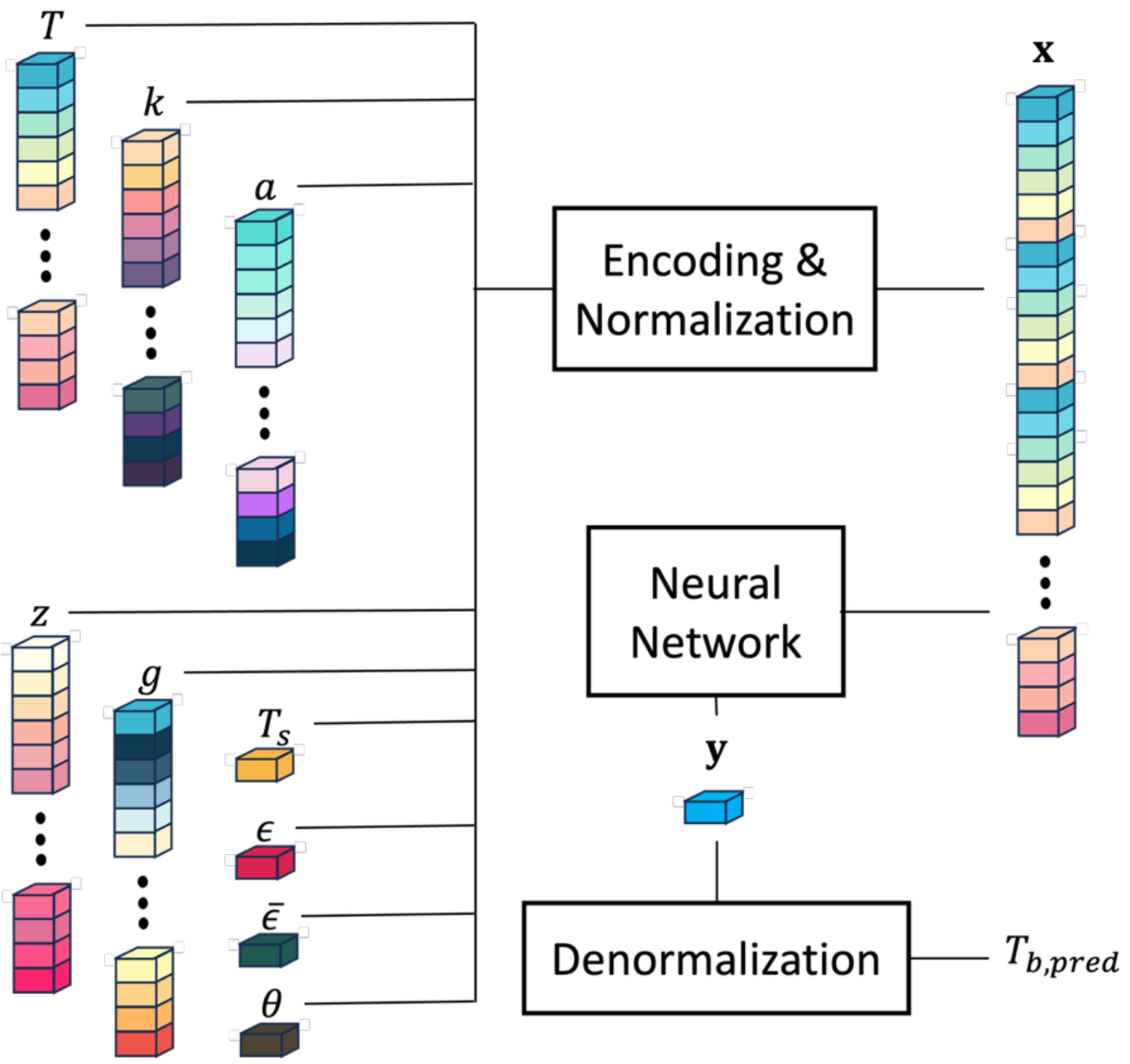

Fig. 1. Emulator architecture.

## III. Results

Training for each instance of the emulator was done using multiple stages, with each stage consisting of 5 epochs. The learning rate was set initially to 0.001 and cosine-annealed within each stage to 0, with warm restarts between stages in order to avoid convergence in local minima (Jones et al., 2026; Loshchilov & Hutter, 2016; Pfreundschuh et al., 2022). Warm restarting was done after the initial stage to half of the previous initial stage rate and held for three stages before being reduced again. This process is illustrated in Fig. 2.

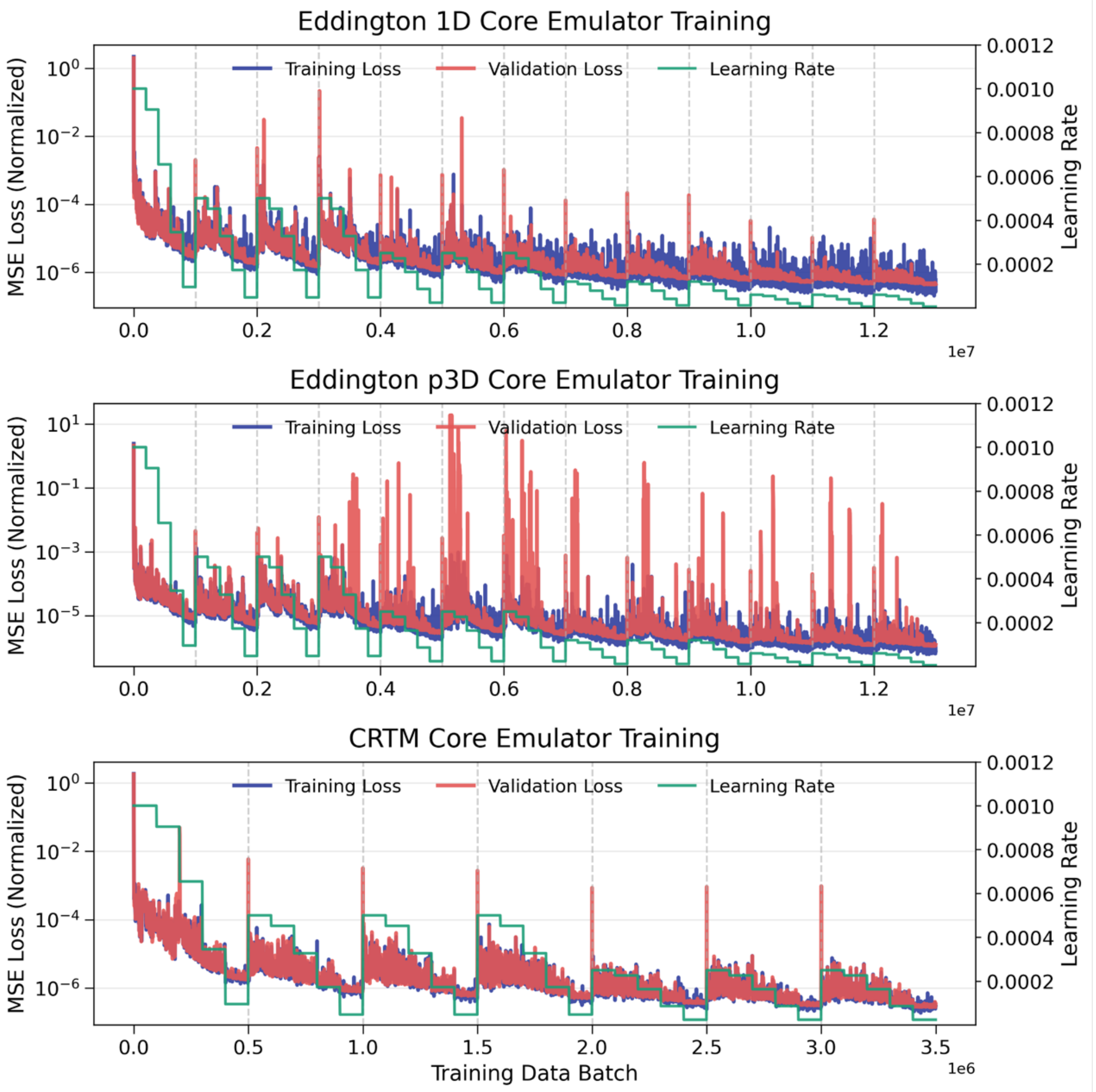

Fig. 2. Emulator training recipe showing the reduction of the loss function throughout training and within each of the 5-epoch training stages. Green lines show the cosine-annealing learning rate (right axis) and warm restarts after each training stage.

As might be expected, more training against the 1DEd and p3DEd models was required than the training against CRTM in order to reduce the loss (taken to be the mean-squared-error between the predicted and true values) against the validation data to within desired parameters. This was again due to the fact that the training data for the 1DEd and p3DEd models were constructed using randomly assigned frequencies from 1 to 300 GHz and view angles from nadir to 65° for each profile, and since CRTM must be configured for specific sensors, only the frequencies and view angle of GMI were used, making prediction a more constrained problem. As Fig. 2. Also illustrates, training against the p3DEd shows a noisier loss reduction after restarting the learning rate, which is due to it having the largest input layer (both upwelling and downwelling columns must be considered separately), and therefore the largest number of predictors to optimize the model against.

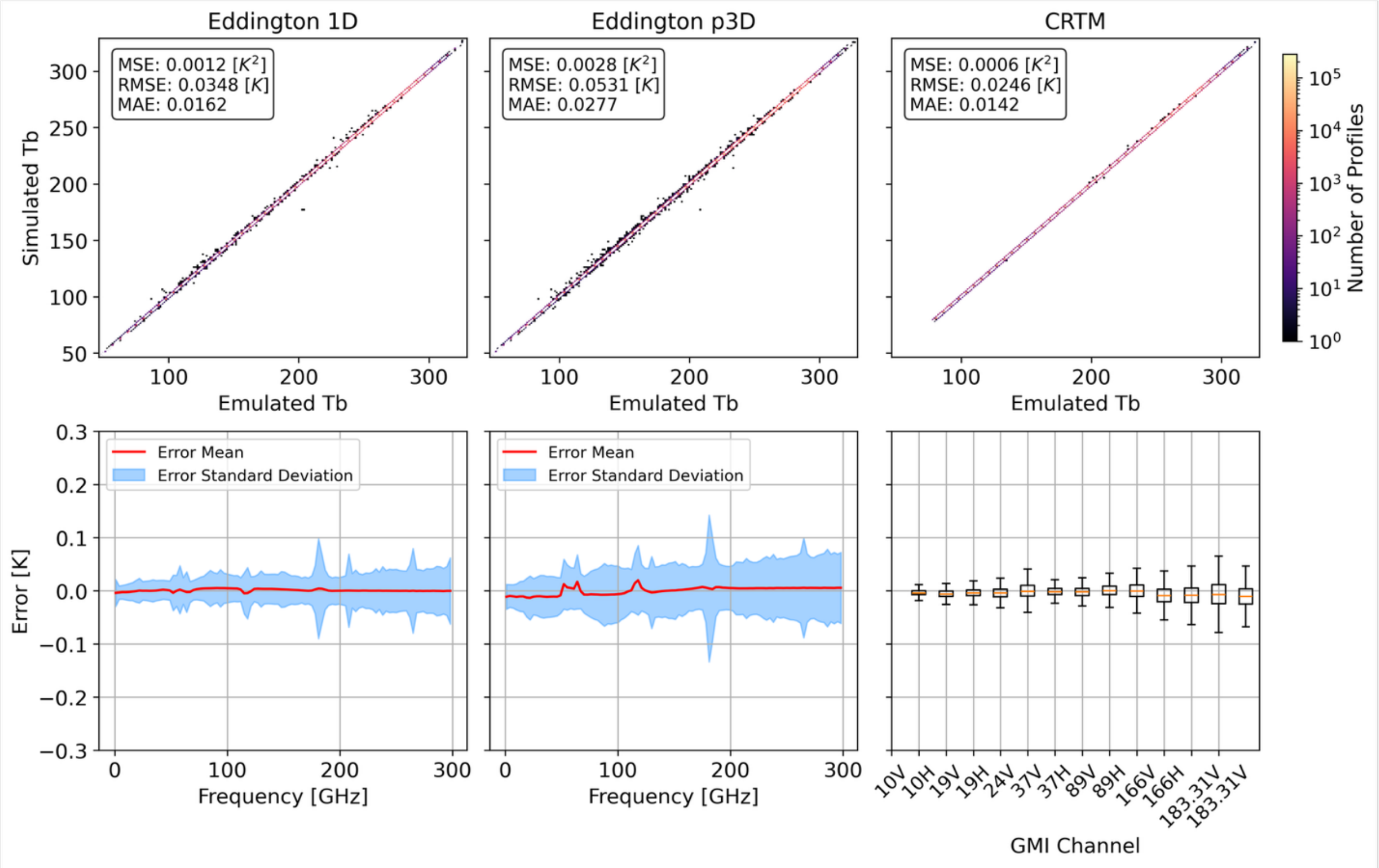

Fig. 3. Performance against the independent test data of the 1DEd (left column), p3DEd (middle column), and CRTM (right column) core emulators. Top row shows scatter plots of simulated vs. emulated brightness temperatures from the test data, and bottom row gives a breakdown of error statistics by frequency or by channel in the case of the CRTM core emulator.

As illustrated in Fig. 3, the error of the emulator with respect to the reference model is very small, and the root-mean-squared error and absolute error are on the order of 0.05 K or less, as well as generally unbiased. Since, for any real sensor, the sensor noise, quantified by the noise-equivalent differential temperature (NEDT), is often on the order of 0.5-1.0 K for most passive microwave imagers and sounders (Berg et al., 2016), any introduced error in the emulation method can generally be considered small for most applications that require the explicit quantification of radiative transfer model uncertainties compared to a physics-based reference model. Additionally, as illustrated by panels d-f, the distribution of errors is a function of scattering and absorption at any given frequency. Errors have the widest distributions, and sometimes small residual biases, near absorption bands due to natural variability, and error distributions generally widen as frequency increases due to more sensitivity of the $T_b$ to scattering.

However, there are a very small number of extreme outliers present in the prediction error against the test data, as seen in panels a-c of Fig. 3. These were found to be related either to excessively heavy scattering at high frequencies throughout the atmospheric column that was very rarely seen in the training data or to excessive bounds violations in the normalization. The latter case can happen when quantities that have a small dynamic range in the training data are linearly normalized using min/max normalization. Since the testing data are normalized using the parameters calculated from the training data, values in the testing data that exceed the minimum or maximum values seen in the training data are linearly scaled outside of the normalization bounds (see eq. 7). This means that small deviations from the maximum or minimum value in predictors with a narrow dynamic range can result in large outliers as seen by the neural network, regardless of the relative importance of the predictor. As an example, some normalized values of upper tropospheric temperature in the testing data were found to be on the order of 10% higher than the values seen in the training data. Again, the excessive normalization bounds violation was not seen in the training data and the sensitivity of machine learning predictions to out-of-sample predictors causes a large bias in the result. However, these cases occur very rarely, and 0.006% and 0.013% of the test data in the 1DEd and p3DEd cases have an absolute error of more than 1 K.

Overall, these results indicate that the emulator is able to perform extremely well against all three reference models for any microphysical and surface parameterization that is within the broad scope of possibilities allowed for in the training data. This is again due to the fact that a large variance of PSD parameters and surface characteristics were represented.

## IV. Implications for Exact 3D Emulation

While the approach to emulate a general core radiative transfer method using FNN architectures has been shown in this paper to be effective for radiative transfer methods that use a limited horizontal structure in one or two dimensions, the question naturally arises whether these models can successfully learn the spatial structure of a full 3D scheme. Approaches like 3D Monte Carlo or 3D discrete ordinate methods are often seen as "exact" methods due to their explicit handling of horizontal and vertical spatial structure in the radiative transfer problem, but their extreme computational cost has relegated them to special studies and theoretical treatments. The promise of a low-cost, machine learning emulator for this type of radiative transfer, is, however, an intriguing prospect with many potential applications. It is obvious that a very large amount of computational power would be required for the development and training of such an emulator, and this kind of computation would be outside the scope of this study. However, a simple experiment has been performed 1) to assess whether the generalized emulation structure is applicable to this type of problem, and 2) to estimate how much computational time and training data would be required to successfully train a general emulator on an exact 3D scheme.

For this experiment, the reference model used was a 3D backward Monte Carlo radiative transfer model (Roberti & Kummerow, 1999), where each scene consists of a horizontally and vertically varying atmosphere with an embedded cloud field, and photons are tracked by releasing them from the sensor location and allowing them to be scattered and absorbed along the path defined by the incidence angle. Photons that are absorbed are considered as having originated that location with its temperature.

The cloud field used for this experiment was a 100 x 100 km precipitation scene over ocean observed by the Global Precipitation Mission Dual Frequency Precipitation Radar (GPM-DPR) on 2019-08-06. The CORRA hydrometeor fields used were the retrieved rain and precipitation ice water content at the native 5-km horizontal resolution of the Ku-band. The vertical structure of the atmosphere was created by averaging and interpolating the 250-m radar bins to the fixed geometric heights from the surface to 18 km used in Pfreundschuh et al. (2023). Ancillary data not able to be retrieved by the radar and radiometer observations, such as the temperature and pressure profile, surface temperature and wind conditions, cloud liquid and ice water, etc. were taken from the Japan Meteorological Agency's Global Analysis (GANAL), provided in the CORRA data.

To create a training dataset that was manageable to store and implement and yet could also provide a preliminary assessment of how well the model can be trained to learn how the calculated $T_b$ changes with the changing spatial structure of cloud fields, the native 5-km resolution CORRA fields were averaged and regridded to various lower resolutions, and then training was attempted using the same training recipe described in Section III. The environmental variables, i.e. the temperature, pressure, and humidity profiles, as well as the surface temperature and wind, were kept at constant values, taken to be the mean of the field. The cloud and hydrometeor columns were then shuffled within the environment to generate the training data. A total of 500,000 random permutations of the cloud field were used in each of the resolution cases as the training, testing, and validation data, and 80%/10%/10% training/testing/validation split was again used.

The computational point (i.e. where the photons enter the scene from space) for this experiment was chosen to be at $(50, 25)$, or at $1/4$ the distance across the y-axis, and entering the top of the atmosphere at a 50° incidence angle to be approximately that of a passive microwave imager. The frequency was chosen to be 89 GHz to get a mixture of both scattering and absorption, and the computational point was selected in order to maximize the probability of interactions within the scene and to minimize the probability of photons escaping out of the scene. Any photons that escaped the scene were considered to be absorbed by the background environment.

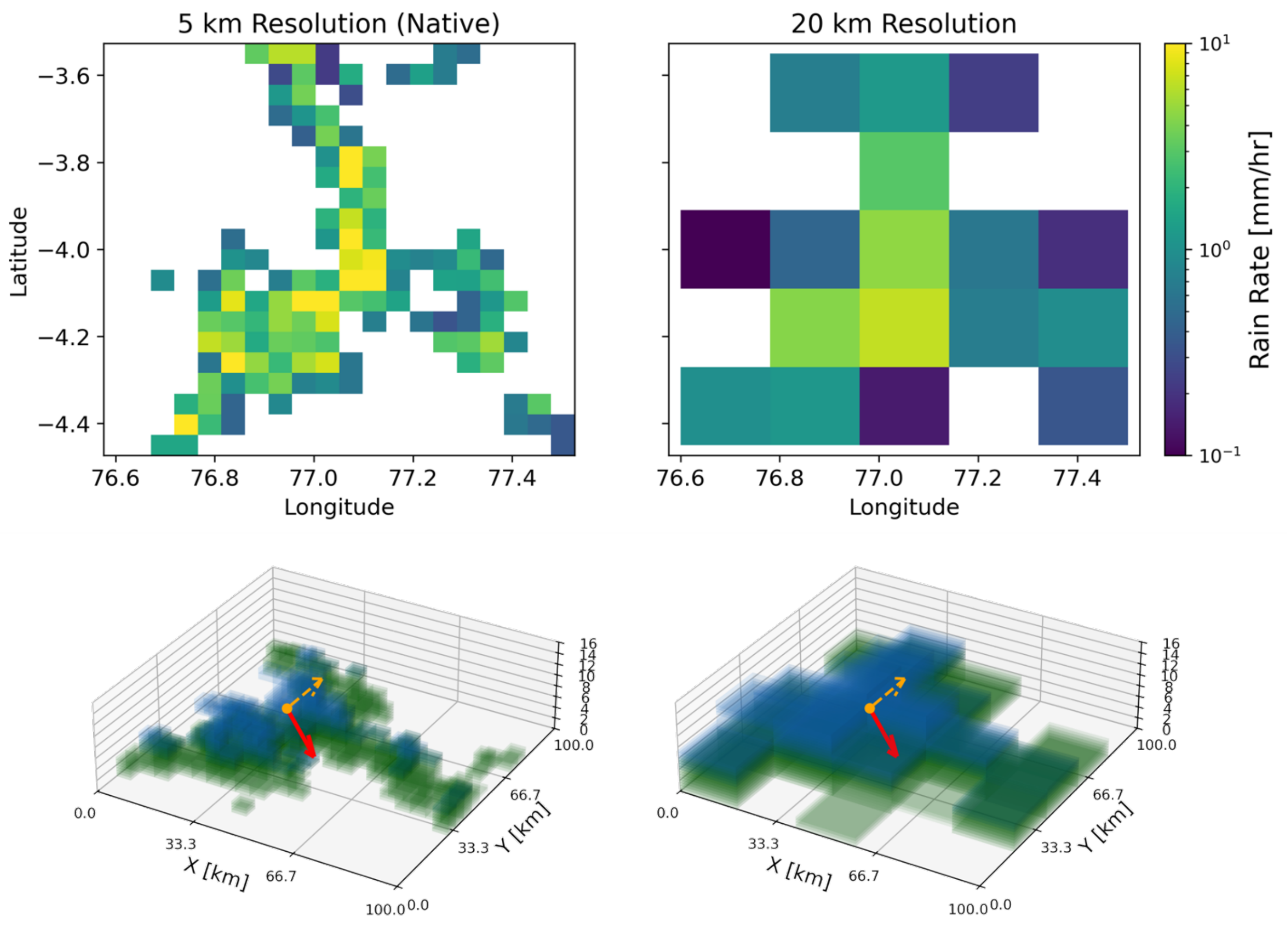

Fig. 4: Observed precipitation scene and retrieved rain rate by the GPM-DPR on 2019-08-06 at native resolution (top right) and regridded to 20 km resolution as an example of one of the lower resolution fields used to create the training data for the 3D Monte Carlo core emulator. Bottom row shows these scenes in 3D form. Green (blue) fields of varying transparency are meant to qualitatively represent increasing (decreasing) precipitation liquid (ice) water content. Orange dot shows point where photons enter the scene, red vector shows direction of incoming photons, and orange dashed vector shows normal to the y-z plane, for visual reference.

The optical properties of each of the cloud columns were calculated assuming, for radiation at 89 GHz, 1) that cloud water only absorbs and does not scatter significantly, 2) that cloud ice is invisible, and 3) that rain and cold-phase precipitation ice are Mie scatterers that are exponentially distributed according to $n(D) = N_0 \exp(-\Lambda D)$, where $N_0$ are chosen to be $8 \times 10^6$ and $3 \times 10^6$ m$^{-4}$ for liquid and ice, respectively. These columns are then randomly shuffled horizontally within the scene, and the resulting permutations of cloud configurations comprise the training, testing, and validation data. It must be noted that, as resolution increases, the number of predictors increases quadratically, and therefore the input layer grows according to this relationship as well.

From the experimental training outlined in Fig. 5, it is clear that the spatial relationships within the cloud field can be correctly recognized by the emulator in their relative importance for scattering and absorption. Even considering the limited scope of the model parameters (i.e. only one frequency and EIA with a constant background), training on shuffled cloud fields with a variety of ice and liquid water contents means the emulator was exposed to many different horizontal and vertical configurations of clouds and was able to learn these interactions successfully.

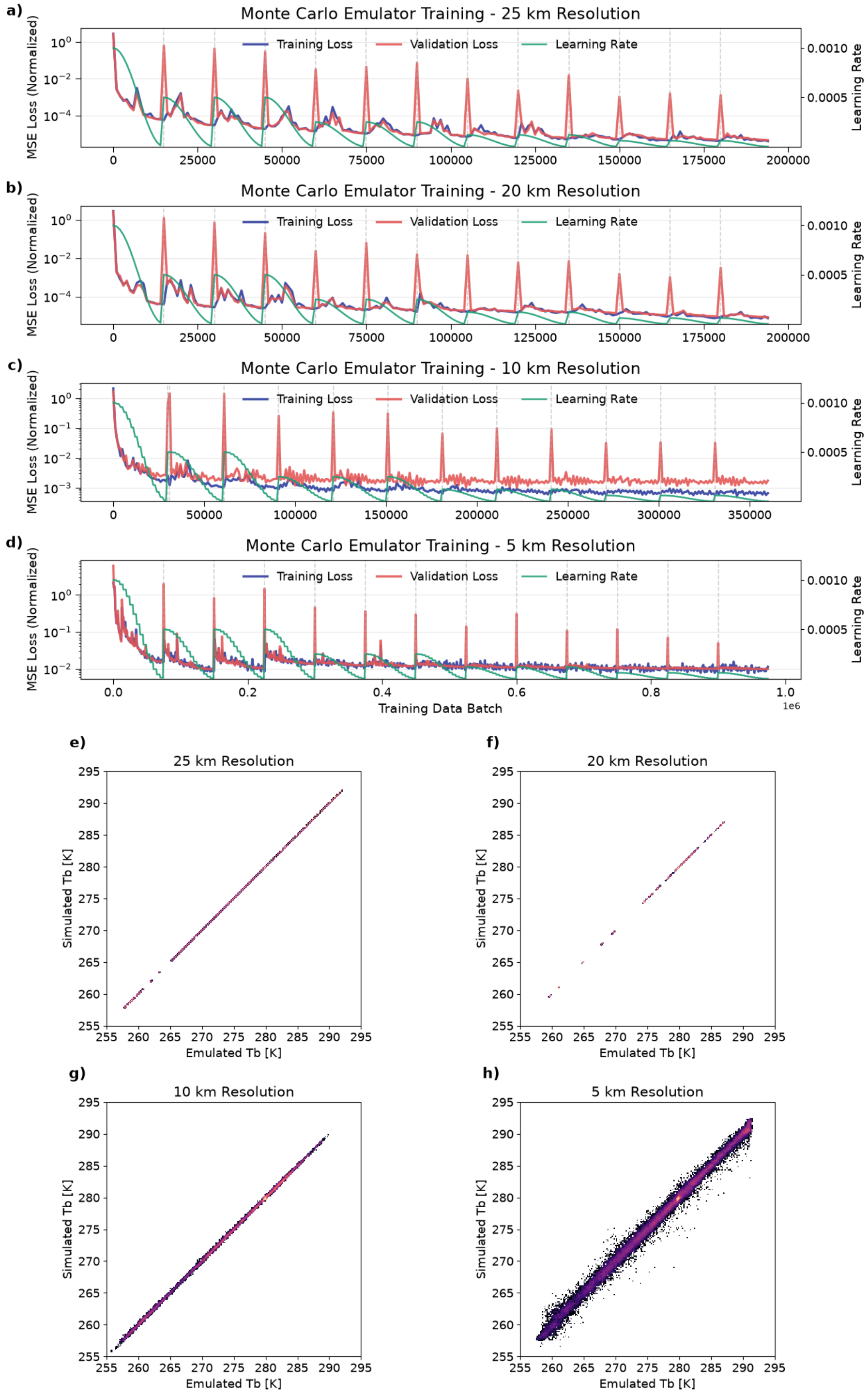

Fig. 5 Experimental training of the full 3D Monte Carlo emulator. Panels a)-d) show loss reduction throughout training. Panels e)-h) show scatterplots of emulator predicted Tb vs. computed Tb by the reference model.

As would be expected, higher resolution also results in a much noisier reduction in loss values, simply due to the much larger number of inputs and tuning of the neurons during training. Again, for a full 3D scene, the number of inputs increases quadratically with the resolution, resulting in an input layer of 33,600 predictors for the 5 km case, and therefore it was necessary to slightly increase the complexity of the network architecture by doubling the size of the first hidden layer from 512 to 1024 neurons in order to achieve a loss value of 0.3 $K^2$ against the test data.

Therefore, we conclude that the emulation of an exact scheme by training on optical properties, such as the 3D Monte Carlo reference model used in this experiment, is possible, though limiting the prediction task by training on only one frequency or domain could make it much more feasible with reasonable computing power. Generalization, as in the case of the 1DEd and p3DEd core emulators described in Section III, would require much more training data and time, as training on real observational or model data would require varying the background environment as well as the cloud field.

From this experiment, we can now estimate the resources needed for creating and training an FNN with this design to emulate full 3D radiative transfer. The first challenge is the creation of the training dataset itself. For good sampling of natural variability, the dataset would likely need to be on the order of 100 million scenes or more, based on the results in Section III. Unlike the experiment performed here, a wide variety of environmental conditions would need to be sampled, and these environmental parameters would need to be added as predictors along with the optical properties. Furthermore, for higher frequencies than the one tested here, it is likely that more moments of the phase function would need to be used for good predictability. Therefore, assuming the following: 1) a single frequency and view angle, 2) 100 x 100 km scenes at 5 km resolution, 3) the set of predictors are $k$, $a$, $g$, and an additional two moments of the phase function for each column, along with environmental predictors consisting of the temperature profile, surface temperature, and emissivity, the dataset would essentially be $2.24124 \times 10^{13}$ bytes at 4-byte precision.

Further challenges include performing the explicit radiative transfer calculations on each scene, which can vary widely with the number of photons used. For this experiment, the number of photons was chosen to be 100,000 based on the desired noise tolerance in the explicit simulations, but this number could be smaller for lower frequencies that do not experience significant scattering. Again, with the assumptions listed above, around 17,400 hours of CPU time would be required to produce the simulations, neglecting the time that it would take to calculate the optical properties for each column.

Finally, training the model would require more than 3,000 hours, composed of the 13 stages with 15 epochs each, based on the average of around 3200 seconds per stage on 10 parallel CPUs for the small training dataset. These again are simple estimations based on extrapolating the time and resources required for the experimental training described in this section, and these estimates could easily be reduced with GPU time and other data storage and training optimizations. While these values all exceed what can realistically be accomplished within a university research group, it is not out of the realm of today's possible should real applications for a fast 3-D radiative transfer routine become necessary.

Summary and Discussion

This study has demonstrated the utility of machine-learning emulators for dramatically increasing the speed and lowering the computational cost of core radiative transfer schemes. The generalized method relies on directly using atmospheric optical properties as predictors rather than the geophysical variables themselves, which allows for more flexibility in the implementation. If the full variability of extinction, scattering, and emission characteristics is captured in the training dataset, the emulator can be trained to be as general as possible and therefore does not have to be retrained if the microphysics, scattering, or emissivity schemes (or lookup tables) are changed. This was done in this study by producing the training data with a variety of particle size distribution (PSD) assumptions by using the normalized gamma form of PSD and allowing the moveable parameter $N_w$ and the ice particle density to be assigned randomly. This, combined with the use of nearly 100 million global ERA5 profiles of atmospheric and surface characteristics, ensured that a wide variety of absorption and scattering inputs were seen by the emulator during training.

The emulator architecture was trained separately to predict the top-of-atmosphere $T_b$ calculated by three separate reference models: a one-dimensional Eddington two-stream model (1DEd), a pseudo-three-dimensional Eddington two-stream model (p3DEd), and the Community Radiative Transfer Model (CRTM). The training data for each model was taken from ERA5 global reanalysis profiles from the 1st and 15th day of each month of 2024, and the validation and testing data were taken from the 1st and 10th day of January, April, July, and October 2025. Frequencies and earth incidence angles in the training data for the 1DEd and p3DEd models were randomly assigned to each profile, removing explicit dependence on frequency and view angles and increasing the utility of such an emulator for coupling with high-resolution models, use in observational studies that require computationally expensive radiative transfer simulations, or to create simulated observations from future or proposed sensors that do not yet exist. For the emulation of the CRTM core method, the view angle of GMI and GMI channel frequencies were randomly assigned to the training data profiles. In all cases, the FNN architecture was able to emulate the core radiative transfer method to within a 0.05 K root-mean-squared error.

A simple experiment was performed to assess if it were possible for the core emulation method to be applied to an exact 3D scheme, with promising results. A backward Monte Carlo radiative transfer model was used to produce a small amount of training data consisting of 100 x 100 km scenes, which were regridded to various resolutions in order to assess the ability of the emulator to learn the relative importance of the location of scatterers and absorbers in a real-world cloud field. Using the optical properties of the scene helps to constrain the prediction, although it is likely that using two or three moments of the phase function would greatly improve model performance at resolutions higher than 10 km. Preliminary training shows that an FNN with a fixed input layer can successfully learn the spatial relationships, though computational resources to generate and hold the training data volumes that would be required for high resolution and to fully train an emulator are outside the scope of this study.

Finally, as Table 3 shows, the main advantage to the use of a core radiative transfer emulator is speed. Each of the table fields list computational time of the core radiative transfer calculations of each reference model used in this study using original Fortran routines executed on a CPU. The GPU-enabled emulator can produce simulated observations in a small fraction of the time. Reductions in computational costs for radiative transfer such as these can open the door for removing the practical limitations of these models in the future for high resolution climate model and sensor constellation studies.

Table 3: Computational speed gained by using core emulator vs. the explicit computations of the reference models. Compute times are listed as wall time in seconds per 100,000 profiles.

| | **1D Eddington** | **CRTM** | **p3D Eddington** | **Monte Carlo** |
|---|---|---|---|---|
| Reference Model (CPU: Intel Xeon Gold 6230) | 3.15 | 14.53 | 6.20 | 49,621.4 |
| Emulator (CPU: Intel Xeon Gold 6230) | 3.54 | 3.17 | 4.43 | 5.86 |
| Emulator (GPU: NVIDIA A100-PCIE) | 0.0019 | 0.0014 | 0.0016 | 0.0014 |

Acknowledgements

This work was supported by the National Aeronautics and Space Administration (NASA) under grant number 80NSSC22K0604 and Future Investigators in NASA Earth and Space Science and Technology (FINESST) under grant number 80NSSC25K0480. No use of generative AI was involved in the writing of this article.

Data Availability

All data used for the research described in this article is publicly available. ECMWF ReAnalysis (ERA5) fields are available from (Copernicus Climate Change Service, Climate Data Store, 2023) at https://doi.org/10.24381/cds.adbb2d47. The emulator described in this article is available as a free software package titled the Microwave Radiative transfer Generalized Emulator (MiRAGE) at https://github.com/SpencerRJones/mirage. Instructions can be found for installing and using the model, as well as retraining the model to emulate other radiative transfer schemes.